\documentclass[preprint,preprintnumbers,amsmath,amssymb]{revtex4}

\usepackage{graphicx}
\usepackage{dcolumn}
\usepackage{bm}
\usepackage{amssymb}
\usepackage{wasysym}

\begin{document}

\preprint{}

\title{Space Charge Effects in Ferroelectric Thin Films}

\author{P. Zubko}
  \email{pz212@cam.ac.uk}
\author{D. J. Jung}%
  \altaffiliation[Also at ]{Advanced Technology Team, Semiconductor R\&D Center, Samsung Electronics Co. LTD,  San\#24, Nongseo-Dong, Giheung-Gu, Yongin-City, Gyunggi-Do, South Korea}
\author{J. F. Scott}
\affiliation{%
Centre of Ferroics, Department of Earth Sciences, University of Cambridge, Cambridge CB2 3EQ, United Kingdom 
}%

\date{\today}

\begin{abstract}
The effects of space charges  on hysteresis loops and field distributions in ferroelectrics have been investigated numerically using the phenomenological Landau-Ginzburg-Devonshire theory. Cases with the ferroelectric fully and partially depleted have been considered. In general, increasing the number of charged impurities results in a lowering of the  polarization and coercive field values. Squarer loops were observed in the partially depleted cases and a method was proposed to identify fully depleted samples experimentally from dielectric and polarization measurements alone. Unusual field distributions found for higher dopant concentrations have some interesting implications for leakage mechanisms and limit the range of validity of usual semiconductor equations for carrier transport.
\end{abstract}

\maketitle

\section{Introduction}
While in bulk they behave like insulators, the description of ferroelectrics as wide-band-gap semiconductors is now  widely used to account for many properties of thin films. The semiconducting state is, in general, thought to be detrimental to ferroelectric properties and thus adaptation of standard semiconductor theory to ferroelectric materials has been the theme of many recent publications. However, despite numerous theoretical efforts, no consensus has yet been reached on how to include the effects of semiconductivity when analysing experimental data. Analytical models tend to make simplifying assumptions which, while providing intuitive insight, may miss out important effects or restrict the range of validity of the model to less interesting cases from the experimental point of view, whereas detailed numerical simulations often assume many parameters that cannot be easily extracted from experiments. A common analytical approach is to include a uniform ferroelectric polarization as two sheets of opposite charge diplaced from the electrodes by some distance assumed to be the thickness of a non-ferroelectric (or ``dead'') layer \cite{Dawber_NDR, Pintil1, Pintil2}. Pintilie \emph{et al.} \cite{Pintil1, Pintil2} have used such a model to interpret their electrical measurements on semiconducting epitaxial single crystal (Pb,Zr)TiO$_3$ (PZT). However, as we will show later,  the assumption of a uniform polarization may become invalid  when space charge levels exceed $\sim 10^{19}$cm$^{-3}$, and  spatial variation of polarization $P(z)$ has to be taken into consideration. 

One way of including a non-homogeneous polarization is in the framework of Landau-Ginzburg-Devonshire (LGD) theory. The free energy density is expanded in powers of $P(z)$ and the gradients $\nabla P(r)$ truncated at an appropriate power; its minimization using the variational principle generally results in a second order differential equation whose solution requires the boundary conditions to be known. Since no agreement has been reached on what these boundary conditions are, they currently provide a playground for theoretical investigations into  various surface effects, but at the same time they have a large influence on the polarization profiles within the bulk of the films, as was shown numerically by Baudry~\cite{Baudry1999, Baudry2005}.

Ignoring the gradient term in the free energy expansion leads to a  polynomial equation, which was investigated analytically by Bratkovsky and Levanyuk~\cite{Bratkovsky}. They mainly looked at the effects of space charges on phase transitions showing a lowering of the transition temperature and suppression of ferroelectricity in the near-electrode layers. However, only cases where the space charge was localised as either one central sheet or two symmetrically placed sheets of opposite charges were considered, and some of the analytic results are only valid for low doping concentrations. A polynomial equation was also addressed numerically by Brennan \cite{Brennan}, who included the effects of polycrystallinity and looked at uniform doping but again only for low doping concentrations.

Finally, we would like to mention the work of Morozovska \cite{Moroz_FE_317, Moroz_PhysicaB,Moroz_JPhysC,Moroz_PhysStatSolidi} who approached the problem of space-charge effects using the Landau-Khalatnikov equation, which has a time-dependent term and therefore allows the simulation of hysteresis loops at various frequencies. Apart from confirming the constriction of the hysteresis loops shown by Baudry and Bratkovsky and Levanyuk, Morozovska \emph{et al.} have shown that 
dielectric losses may be responsible
the tilting of the loops, as well as reported other effects such as the appearance of double hysteresis loops. Due to the complexity of the equations, however, the spatial variations of polarization and electric field were not investigated and only volume-averaged values were reported.

None of the studies above (apart from \cite{Brennan}) have discussed in any detail the electric fields inside ferroelectric samples with space charge, and to date researchers of leakage currents in ferroelectrics are divided into those who interpret their data assuming a constant field, compatible with fully depleted semiconductors or insulators, and those who use the standard results for partially depleted semiconductors where the field drops linearly across a depletion layer to a negligible value in the bulk.

We have focused on more realistic space charge distributions and  simulated the electric field distributions and ferroelectric hysteresis loops in fully and partially depleted ferroelectric films using parameters comparable to those extracted experimentally for a high quality 144nm thick PbZr$_{0.4}$Ti$_{0.6}$O$_3$ thin film as an example. We show a decrease in coercive field and remanent polarization, $P_r$, values with increased dopant levels as observed in previous works, but also some unusual field profiles and discuss their significance for leakage mechanisms and coercive fields in ferroelectric thin films. Last, we discuss the main problem of the thermodynamic modelling approach, namely its overestimation of the coercive field $E_C$  if bulk parameters are used. It will be discussed how the experimentally measured values of $E_C$ can be reconciled with the phenomenological approach provided space charge is taken into account. 

\section{Model and assumptions}
\subsection*{Landau-Ginzburg-Devonshire theory}
In LGD theory, the free energy $F$ is composed of a bulk term $F_b$, given as a power series expansion in $P$, and a surface term $F_S$. The energy of a dipole $\textbf{p}$ in a field $\textbf{E}$ is $-\textbf{E}\cdot\textbf{p}$; however, when considering the interaction of the local polariasation $P(z)$ with the local field $E(z)$ one has to be careful not to include the self-field, hence the factor of a half in front of the depolarization field ($E_{dep}$) term in equation~(\ref{FreeEnergyBulk:OK}) (where $S$ is the cross-sectional area, $E_{r}$ is the electric field excluding the depolarization term and $\alpha$, $\beta$ and $\kappa$ are the usual Landau coefficients; $\lambda$ is the so-called extrapolation length \cite{KretschmerBinder}; a second order phase transition has been assumed).
\begin{equation}
\frac{F_{b}}{S}=\int_0^Ldz\;\;\frac{\alpha}{2}P^2(z)+
\frac{\beta}{4}P^4(z)+\frac{\kappa}{2}\left(\frac{dP(z)}{dz}\right)^2
-E_r(z)P(z)-\frac{1}{2}E_{dep}(z)P(z)
\label{FreeEnergyBulk:OK}
\end{equation}

\begin{equation}
\frac{F_S}{S}=\frac{\kappa}{2\lambda}\left(P(0^+)^2+P(d^-)^2\right)
\end{equation}

\begin{equation}
F=F_b+F_S
\end{equation}

Only terms up to fourth power in $P(z)$ have been retained for the calculations in this paper, however, higher powers can be added without any additional difficulty when modelling first order phase transitions or bulk tetragonal PZT, which has high polarization values thus not allowing the $P^6(z)$ term to be neglected.

Using the variational method to minimize the free energy yields a non-linear second order differential equation
\begin{equation}
\alpha P(z)+\beta P^3(z)-\kappa \frac{d^2P(z)}{dz^2}=E_r(z)+\frac{\bar{P}-P(z)}{\epsilon_0}
\label{SecondOrderDE:OK}
\end{equation}
with boundary conditions determined by the surface term $F_S$
\begin{equation}
\left(\frac{dP}{dz}\right)_{z=0}=\frac{P(0^+)}{\lambda};
\left(\frac{dP}{dz}\right)_{z=d}=-\frac{P(d^-)}{\lambda}.
\end{equation}
An approximate analytic solution to equation~(\ref{SecondOrderDE:OK}) was obtained by Glinchuk and Morozovska~\cite{Glinchuk} for the case of an ideal insulating ferroelectric with no space charge. Baudry~\cite{Baudry1999, Baudry2005} attempted to include the space charge effects by solving the equation numerically and found an interesting interplay between the surface effects due to the imposed boundary conditions and the electric field due to the space charge. However,  due to the unspecified physical nature of the surface effect the calculated field and polarization profiles in his case seem to be inconsistent with the electrostatic equations used. The main problem in solving equation~(\ref{SecondOrderDE:OK}) is that the parameter $\lambda$ is unknown. Recent \emph{ab initio} calculations for epitaxial KNbO$_3$ with SrRuO$_3$ and Pt electrodes by Duan \emph{et al.} \cite{Duan} have shown that what happens near the surface is highly dependent on the type of atomic bonding between the ferroelectric and the metal, with both reduction and enhancement of local polarization  possible.

In this paper we are only interested in the effects of space charge and will therefore ignore what happens at the interface, especially since so little is known about the boundary conditions that should be used there. This assumption is most realistic for thicker films. We also neglect the gradient term in the bulk, which will be fully justified later. 

\subsection*{Electric fields}
\begin{figure}[h]
\centering
\scalebox{0.5}{\includegraphics{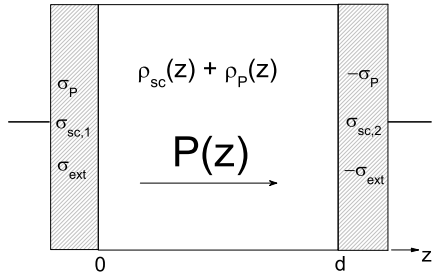}}
\caption{Volume charge densities ($\rho_{sc}$ and $\rho_{P}$) inside the ferroelectric and surface charge densities ($\sigma_{ext}$, $\sigma_{sc}$ and $\sigma_{P}$) on the electrodes assumed in the model.}
\label{Fig_Setup:OK}
\end{figure}

The sample is assumed to be a uniformly doped monodomain single crystal ferroelectric of thickness $d$ with symmetrical and perfectly metallic electrodes. The sum of the charges inside the sample and on the electrodes must be zero to have zero field in the electrodes. We have chosen to separate the charge distributions as illustrated in figure~\ref{Fig_Setup:OK}. It can be shown \cite{KretschmerBinder} that the polarization $P(z)$ inside the film  will induce a charge density $\sigma_{P}$ on the electrodes equal to the average polarization $\bar{P}$ 
\[\sigma_{P}=\bar{P}=\frac{1}{d}\int_0^dP(z)dz.\]
For a partially depleted sample $\sigma_{sc,1}$ and $\sigma_{sc,2}$ will compensate the charges in the two depletion layers. In the case of full depletion at zero applied voltage $\sigma_{sc,1}=\sigma_{sc,2}=1/2\times$total space charge per unit area in the sample; $\sigma_{ext}$ and $-\sigma_{ext}$ provide the external field.

The electric field at any point in the sample is given  by
\begin{eqnarray}
\epsilon_0 E(z)&=&D(z)-P(z)\nonumber\\
&=& \sigma_{ext}+\sigma_{P}+\sigma_{sc}+\int_0^z\rho_{sc}dz-P(z)\nonumber\\
&=& \sigma_{ext}+\underbrace{\sigma_{sc}+qN_Dz}+\underbrace{\bar{P}-P(z)}\nonumber\\
&\equiv& \epsilon_0 E_{ext}\;\;+ \;D_{sc}(z) \;+\,\,\,\epsilon_0 E_{dep}(z)
\end{eqnarray}
where $D(z)$ is the electric displacement and $E_r(z)\equiv E_{ext}+D_{sc}(z)/\epsilon_0$. Inserting this into equation~(\ref{SecondOrderDE:OK}) and neglecting the gradient term leads to a third order polynomial 
\begin{equation}
\alpha P(z)+\beta P^3(z) =E_{ext}+\frac{D_{sc}(z)}{\epsilon_0}+\frac{\bar{P}-P(z)}{\epsilon_0}
\label{thirdorder:OK}
\end{equation}
 which, however, does not have a simple analytic solution due to the $\bar{P}$ term. This equation is the same as  previously investigated by Bratkovsky and Levanyuk for the simple cases of a single cetral sheet of charge and two such sheets at symmetric planes. 

\subsection*{Strain effects}
In-plane strain caused by the mismatch between the lattice parameters of the ferroelectric and the substrate can either enhance or reduce the polarization of the ferroelectric compared to the bulk value depending on its sign.
Previous works \cite{Pertsev, Catalan} show that the effect of strain is to simply rescale the  local Landau coefficients $\alpha$ and $\beta$, thus if we assume that the effective coefficients (i.e. those that would be extracted experimentally  from the temperature dependence of the dielectric susceptibility and spontaneous polarization) are also simply rescaled, then our model should be valid  for strained films too.

\section{Simulation results}
\subsection{Full depletion case}
In the case of full depletion, the fields entering equation~(\ref{thirdorder:OK}) are given by
\[E_{ext}
\equiv\frac{\sigma_{ext}}{\epsilon_0}=\frac{V}{d},\]
\[D_{sc}(z)=-qN_D(\frac{d}{2}-z),\]
assuming $\rho_{sc}(z)=qN_D$, as would be in the case of an $n$-type material; however, the model is equally valid for $p$-type samples with a simple sign reversal of the appropriate terms. Equation~(\ref{thirdorder:OK}) was then discretized and solved numerically for various dopant concentrations $N_D$. The simulations were carried out as follows: an initial guess was made at the polarization profile, $\bar{P}$ was computed and equation~(\ref{thirdorder:OK}) was solved to obtain the new polarization profile and thus a new $\bar{P}$. The procedure was repeated until the convergence criterion ($\Delta P<10^{-11}$C/m$^2$ for that iteration) was satisfied. The applied voltage was computed by integrating the total field in the sample.  The following Landau coefficients were used in the simulation for the results to be comparable to our experimental data discussed later: $\alpha=-6.36\times 10^7$ mF$^{-1}$ and $\beta=1.38\times 10^9$ m$^5$C$^{-2}$F$^{-1}$ (giving a $P_s\equiv \sqrt{-\alpha/\beta}$ of 0.215 Cm$^{-2}$); $d$ was assumed to be 150nm.
\begin{figure}[h]
\centering
\scalebox{0.7}{\includegraphics{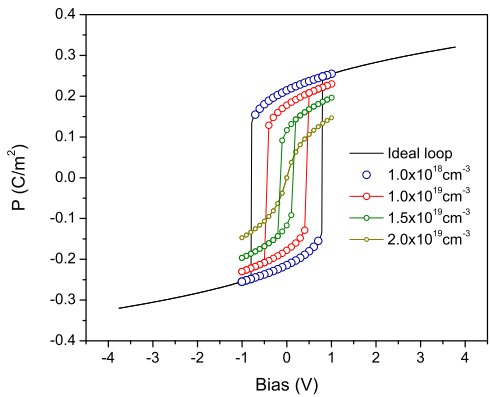}}
\caption{Effect of space charge on ferroelectric hysteresis loop for a fully depleted sample. A reduction from ``ideal'' to zero  is observed in $P_r$ and $E_c$ over a very narrow range of $N_D$.}
\label{FullDepletionLoops:OK}
\end{figure}

Figure~\ref{FullDepletionLoops:OK} shows that for concentrations below about $10^{18}$cm$^{-3}$ the hysteresis loops remain unchanged from the ideal case of uniform polarization. At  $10^{19}$cm$^{-3}$ a clear shrinking of the loop is observed and doubling this concentration leads to an almost total vanishing of $P_r$. So within less than an order of magnitude in $N_D$ the shape of the hysteresis loop changes from almost ``ideal'' to a simple curve.
\begin{figure}[h] 
\centering
\scalebox{0.6}{\includegraphics{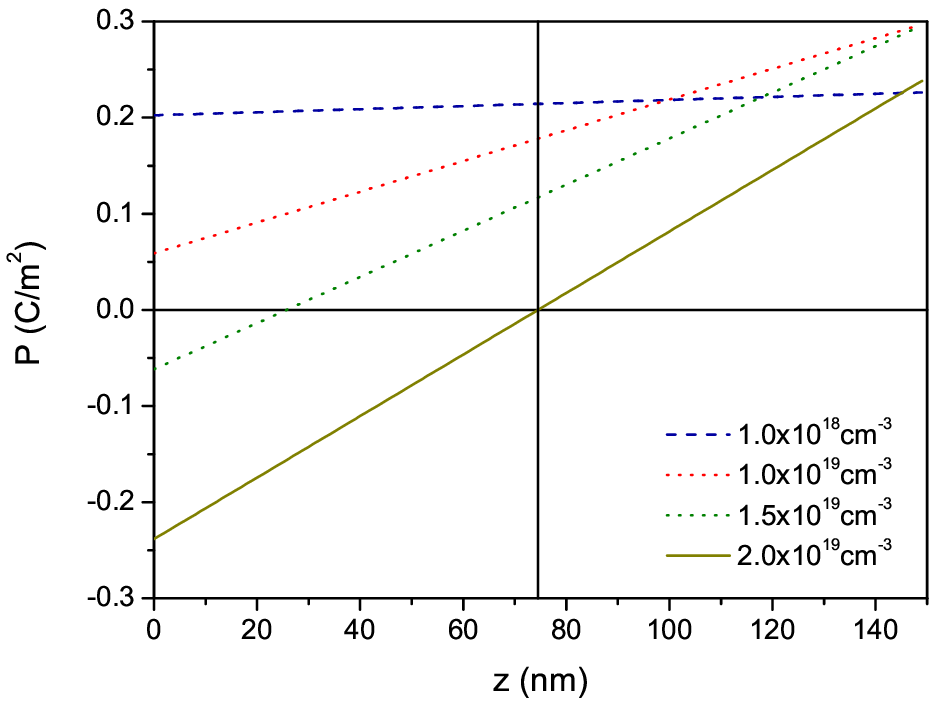}}
\scalebox{0.6}{\includegraphics{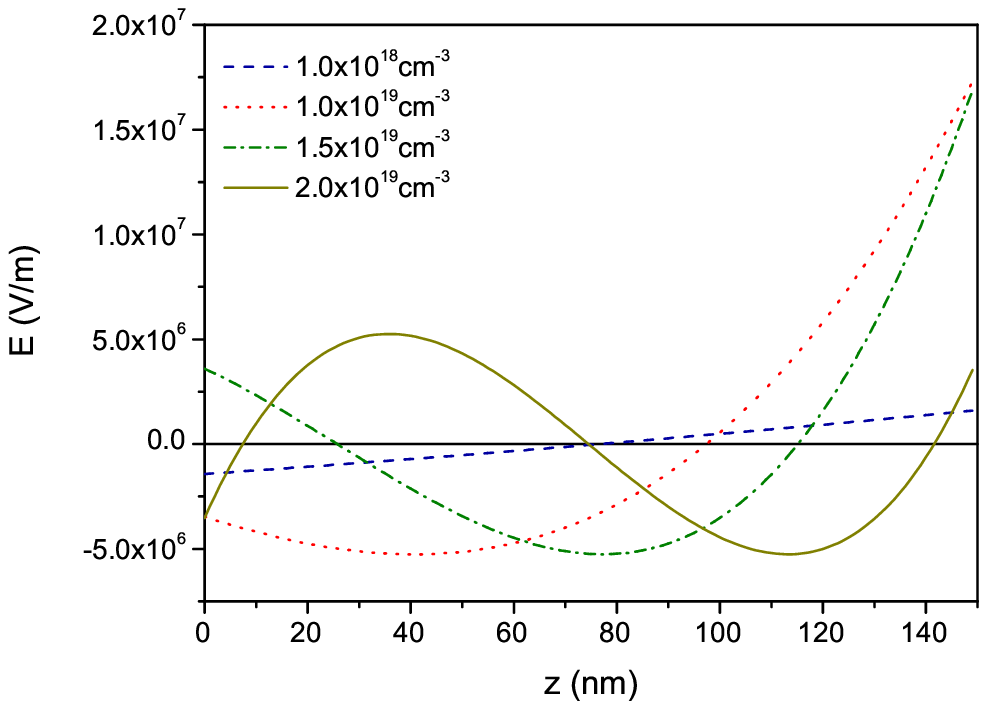}}
\caption{Spatial variation of (a) the local polarization $P$ and (b) the total electric field $E(z)=E_{sc}(z)-(\bar{P}-P(z))/\epsilon_0$ inside the film at zero $E_{ext}$. Note, although $P(z)$ is linear to a very good approximation, it is not \emph{exactly} linear as can be seen from the curvature of the total field; this, of course, is no surprise, as equation~(\ref{thirdorder:OK}) contains a cubic term.}
\label{PandField:OK}
\end{figure}

To understand these results let's look at the internal field distributions (figure~\ref{PandField:OK}). The total field is given by
\begin{equation}
E(z)=\alpha P(z)+\beta P^3(z).
\label{Cubic1:OK}
\end{equation}
Due to the high dielectric constant of ferroelectrics, to a very good approximation $D(z)=P(z)$. From Poisson's equation
\begin{equation}
D(z)=\bar{P}+D_{sc}(z)+\frac{V}{d}\epsilon_0.
\label{Displacement:OK}
\end{equation}
 The last term is generally much lower than either the first or the second for the usual applied fields and we will neglect it for simplicity. For low space charge concentrations the displacement is approximately constant throughout the film and therefore so is the polarization, perturbed only slightly by the linear $D_{sc}(z)$ i.e. $P(z)=\bar{P}+D_{sc}(z)\approx P_s+D_{sc}(z)$. Inserting this into~(\ref{Cubic1:OK}) and neglecting terms of order $D_{sc}^2(z)$ and higher we obtain 
\begin{eqnarray}
E(z)&=&(\alpha +3\beta P_s^2)D_{sc}(z)\nonumber\\
&=&-2\alpha \cdot D_{sc}(z) \nonumber \\
&=& -2\alpha \cdot qN_D(d/2-z)\nonumber
\end{eqnarray}
i.e. a linear internal field profile just as in the case of a semiconductor with $(\epsilon\epsilon_0)^{-1}=-2\alpha$.

On the other hand, when the space charge dominates (\ref{Displacement:OK}), the polarization $P(z)\approx D(z)\approx D_{sc}(z)$ is approximately linear and antisymmetric about the midpoint of the film, averaging to zero. The total field given by (\ref{Cubic1:OK})  is now
\begin{equation}
E(z)\approx\alpha D_{sc}(z)+\beta D_{sc}^3(z)
\end{equation} 
and is a cubic in $z$ as can be seen in figure~\ref{PandField:OK}. This is very different from the usual field distribution is a semiconductor with space charge. 

\subsection{Partial depletion}

For a semiconducting film with $N_D=2\times 10^{19}$cm$^{-3}$ to be fully depleted at zero applied bias it must have a built in voltage $V_{bi}=qd^2N_D/(8\epsilon_f\epsilon_0)\approx 1.2$V (assuming a uniform field independent dielectric constant $\epsilon_f\epsilon_0=1/2\alpha=888\epsilon_0$), whereas the expected values are closer to a few$\times 0.1$V \cite{Pintil2, JimsBook} thus for high  dopant concentrations the films might be partially depleted.

We will again assume a uniform density of fully ionised donor impurities. In the usual abrupt depletion approximation any region with a non-zero field is completely depleted of free charges and  has positive space charge of volume density $qN_D$. In the regions  free of space charge, the total field is zero 
\[E(z)=E_{ext}-\frac{1}{\epsilon_0}(\bar{P}-P(z))=0\]
and\[P(z)=P_s=\sqrt{\frac{-\alpha}{\beta}},\]
giving
\[E_{ext}=\frac{1}{\epsilon_0}(\bar{P}-P_s).\]
For simplicity we assume that all the voltage is dropped across the reverse biased depletion region of width $w=w(V)$; then $D_{sc}(z)=\pm qN_D(w-z)$ within the depletion region. This assumption is true for $V_{bi}=0$ and is a good approximation for all but very low applied biases when $V_{bi}\neq 0$.

In the simulation, $D_{sc}(z)$ is set by choosing the depletion width $w$ and the corresponding potential difference between the two electrodes is again calculated by integrating the total internal field once the calculation converges and $P(z)$ is found. The results are very interesting.

\begin{figure}[h] 
\centering
\scalebox{0.4}{\includegraphics{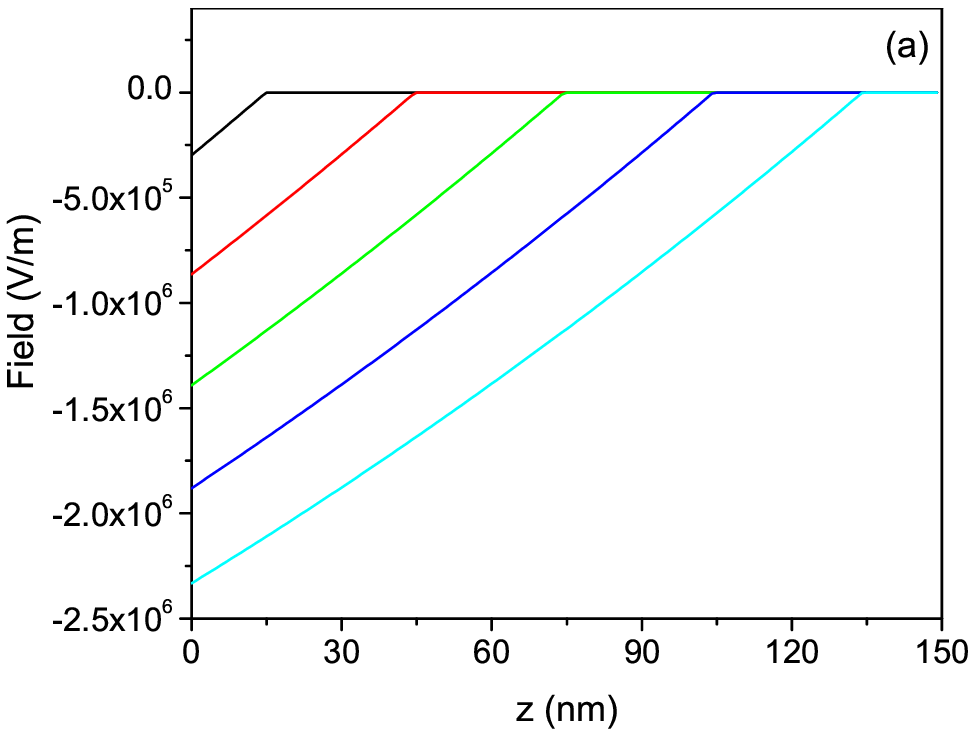}
               \includegraphics{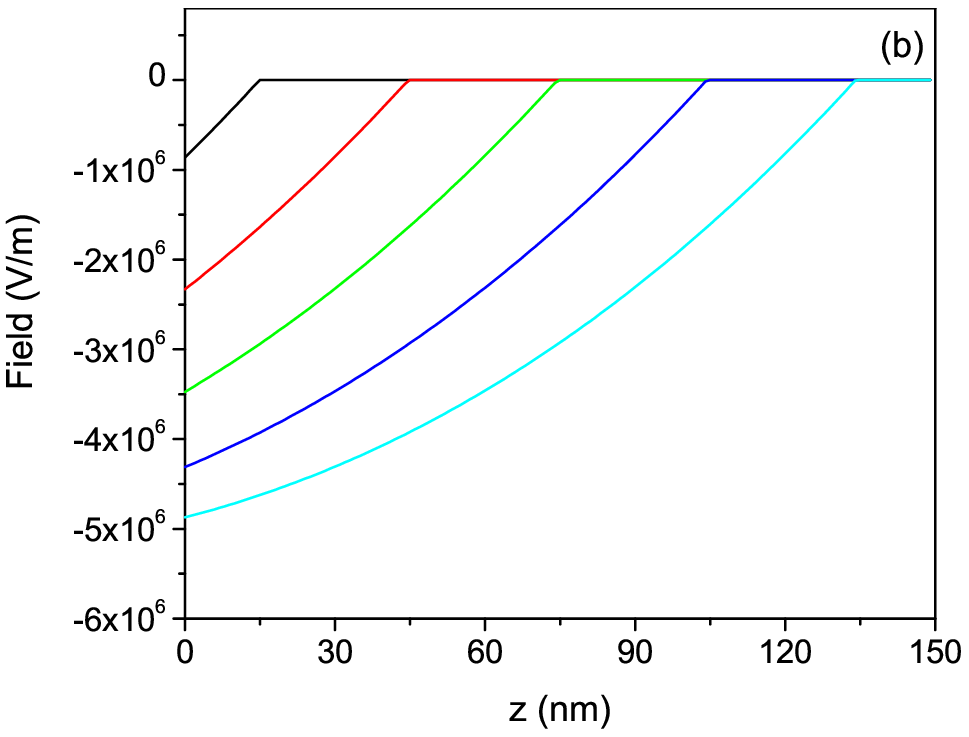}}
\scalebox{0.4}{\includegraphics{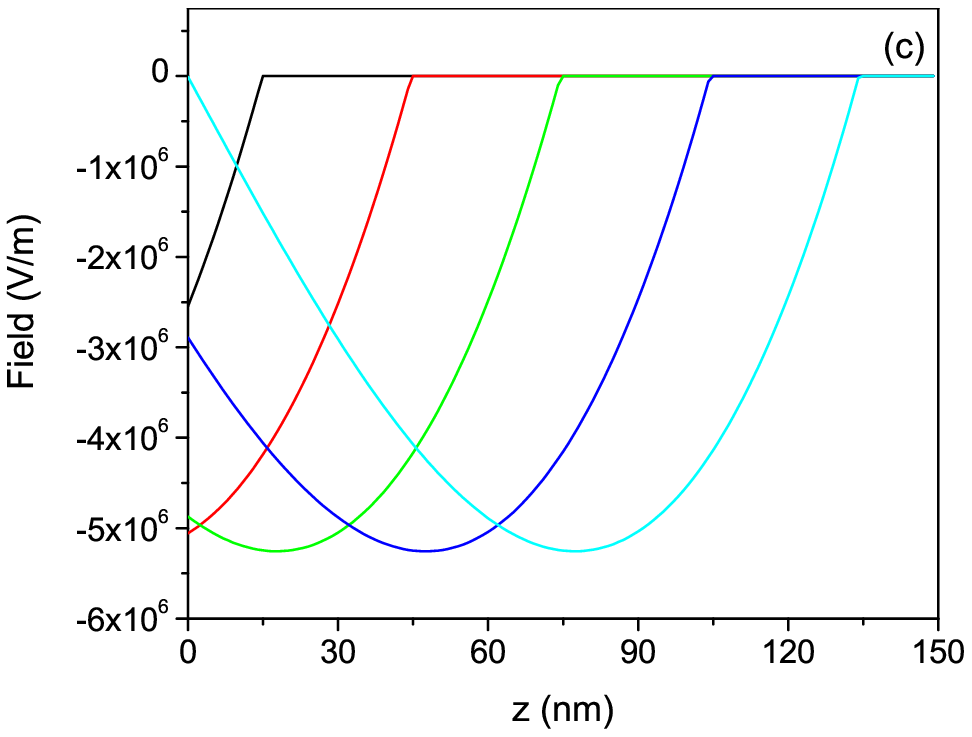}
               \includegraphics{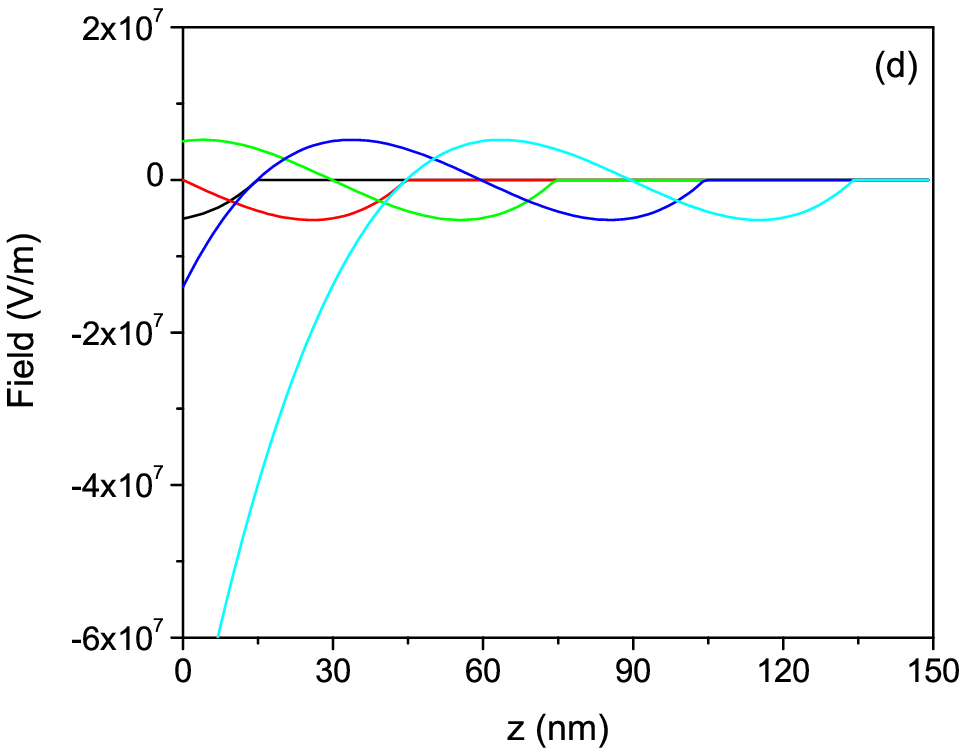}}
\caption{Field profiles inside the 150nm film for different extents of depletion (i.e. at different biases) for four dopant concentrations:  $N_D=10^{18}$, $3\times 10^{18}$, $10^{19}$ and $3\times 10^{19}$cm$^{-3}$.}
\label{FieldPartDepl:OK}
\end{figure}

We start with the sample poled positively (i.e. we force $P(z)=+P_s$ in the space charge free region). Figure~\ref{FieldPartDepl:OK} shows that for small $N_D$ the field increases approximately linearly with $z$ as usual in the  case of semiconductors. The ratio  $D_{sc}(z)/\epsilon_0E(z)$ tends to $\epsilon$ with decreasing dopant concentration as expected. At larger $N_D$, nonlinearities in $E(z)$ due to the cubic term in equation~(\ref{thirdorder:OK}) become obvious, causing an upward bending of the total field near the electrode. This  has great implications for current-voltage characteristics. While the bias is increasing, the field near the interface is reduced and thus the usual field induced potential barrier lowering works in the opposite way -- the barrier is raised, possibly leading to negative differential resistivity. This should be explored further.

 At still higher concentrations (figure~\ref{FieldPartDepl:OK}(c) and (d)) a point is reached when the field at the interface changes sign before full depletion is reached. The bias cannot be increased beyond this point without switching of the polarization. Changing $P(z)$ to $-P_s$ in the undepleted regions gives us the rest of the negative half of the polarization loop. When $N_D\approx 3\times 10^{19}$cm$^{-3}$ a second sign reversal is possible (figure~\ref{FieldPartDepl:OK}d).

\begin{figure}[h] 
\centering
\scalebox{0.7}{\includegraphics{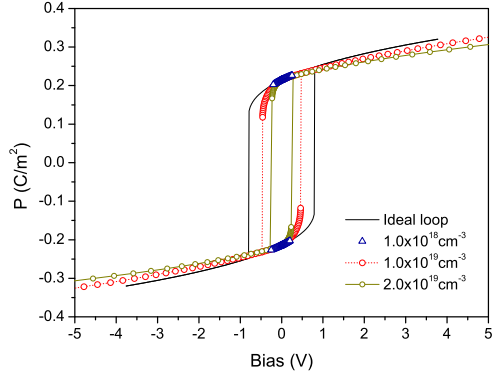}}
\caption{$P$-$V$ loops for partially depleted samples between $\pm$5V. The $N_D=1\times 10^{18}$cm$^{-3}$ loop is only shown in the range where the sample is partially depleted; outside this region the sample is fully depleted and is shown in figure~\ref{FullDepletionLoops:OK}.}
\label{PartDeplLoops:OK}
\end{figure}

The hysteresis loops for three doping concentrations are shown in figure~\ref{PartDeplLoops:OK} between $-5$ and 5V. For the lowest concentration shown full depletion occurs at $\pm 0.25$V and the \emph{P-V} curve shows a good  overlap with the ``ideal'' loop. Increasing the doping leads to a reduction in the coercive voltage and the non-zero field polarization. However, $P_r$ remains the same as there is still no internal field at zero bias for $V_{bi}=0$. In a way, the ferroelectric properties are improved with lower coercive fields, squarer loops and no reduction in the remanent polarization.

\begin{figure}[h]
\centering
\scalebox{0.63}{\includegraphics{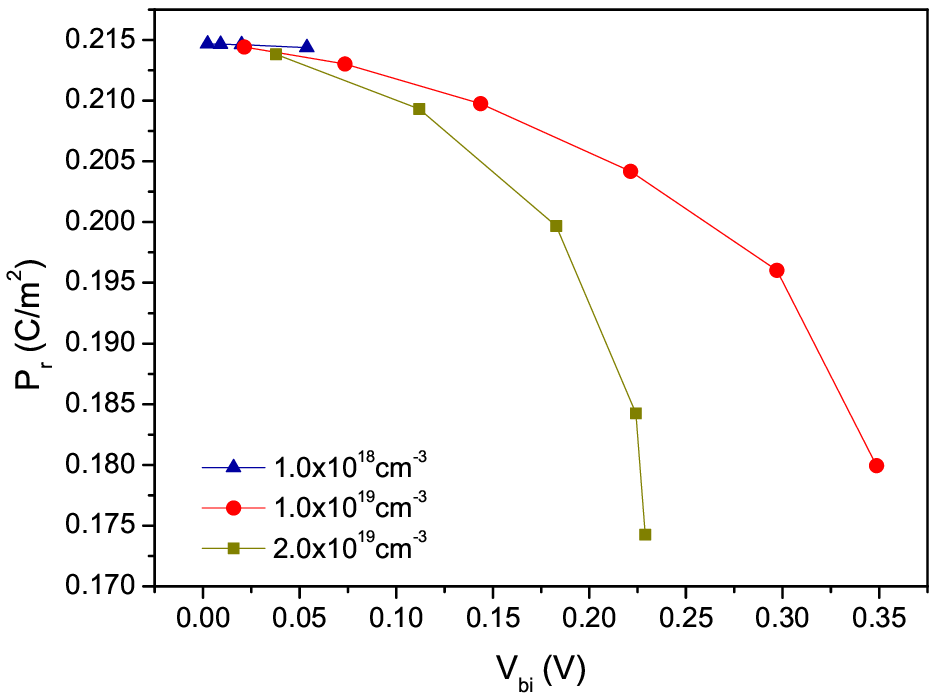}}
\scalebox{0.63}{\includegraphics{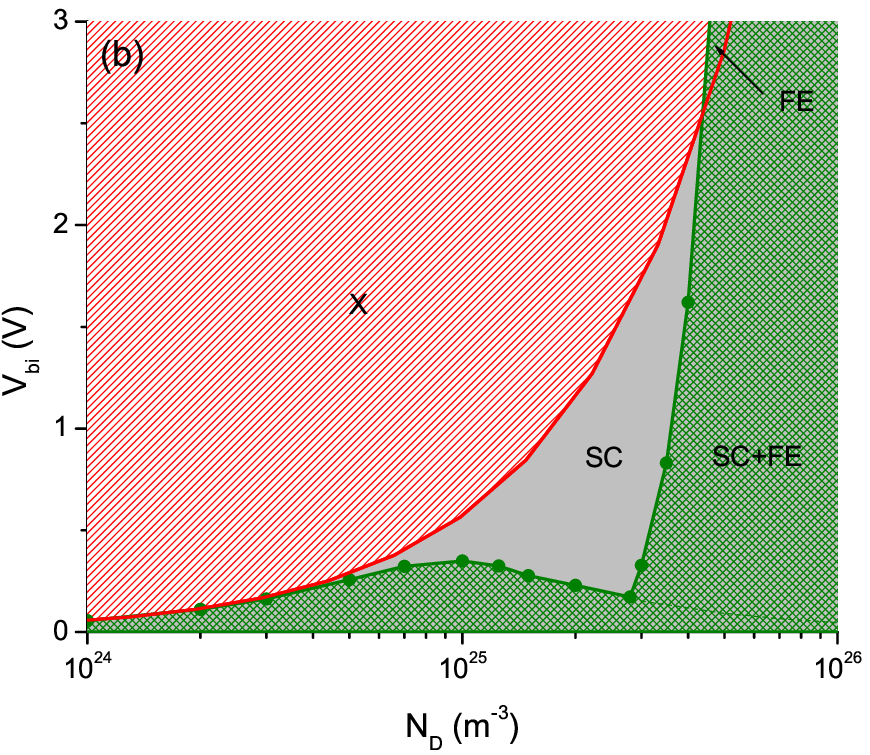}}
\caption{(a) Variation of remanent polarization with built-in bias. (b) Built-in biases which can be supported without full depletion as a function of $N_D$ for a ferroelectric and a semiconductor with $\epsilon\epsilon_0=-1/2\alpha$. Region X is not allowed for either, region SC is only allowed for the semiconductor, FE  for ferroelectric and SC+FE for both.}
\label{PrvsVbi:OK}
\end{figure}

For finite $V_{bi}$, depletion regions  form at both electrodes at zero bias, introducing space charge into the sample and thus lowering the remanent polarization in a similar way as in the fully depleted case. Figure~\ref{PrvsVbi:OK}(a) shows how $V_{bi}$ affects $P_r$. As $N_D$ increases higher voltage drops can be supported before full depletion is reached, as is usual in semiconductors, but within a certain range of $N_D$ the electric fields in the deletion regions start to change sign and the maximum  voltage drop that can be supported decreases. Part (b) of the figure compares the built-in voltages that can be supported by the space charge regions in a ferroelectric and a semiconductor of the same dielectric constant. The maximum around $10^{25}$m$^{-3}$ corresponds to the point at which the electric field at the interface changes sign for the first time due to the cubic term, whereas the minimum around $3\times 10^{25}$m$^{-3}$ corresponds to a second reversal of sign. 

Stricly speaking, at very low biases, half of the applied voltage is dropped across each depletion region in a semiconductor. As the applied bias increases, the fraction dropped across the forward biased depletion region $f_f$ rapidly drops to zero. The exact fraction of the voltage dropped across each depletion width can be determined by equating the currents flowing through both regions, but this involves assuming a specific leakage mechanism. Usually, however, $f_f$ is only significant at \emph{very} low biases, thus we can simulate hysteresis loops for samples with non-zero $V_{bi}$ assuming all the bias is applied across the reverse biased junction. In general, the higher the built-in bias the more space charge there is in the film and the more the loops will tend to those of fully depleted films with same dopant concentration.

\subsection{Comparison with experiments}

For an ideal bulk ferroelectric, whose free energy may be truncated at the $P^4(z)$ term, the $\alpha$ and $\beta$ coeffcients can be found experimentally by measuring the zero field dielectric constant and remanent polarization $P_r$ and using the following relations

\begin{equation}
\frac{1}{\epsilon\epsilon_0}\approx\frac{1}{\chi}=-2\alpha=\frac{2(T_c-T)}{\epsilon_0 C}
\end{equation}
\begin{equation}
P_r^2=P_s^2=-\frac{\alpha}{\beta}=\frac{(T_c-T)}{\beta\epsilon_0 C}
\end{equation}
where $C$ is the Curie constant. Parasitic series capacintaces which reduce the effective measured dielectric constant can be eliminated by looking at the temperature dependence of $\epsilon$. Provided it is temperature independent, the parasitic contribution will not affect the slope of the $1/\chi$ vs. $T$ plot which gives the Curie constant and will  change only the intercept. 
\begin{figure}[h]
\centering
\scalebox{0.7}{\includegraphics{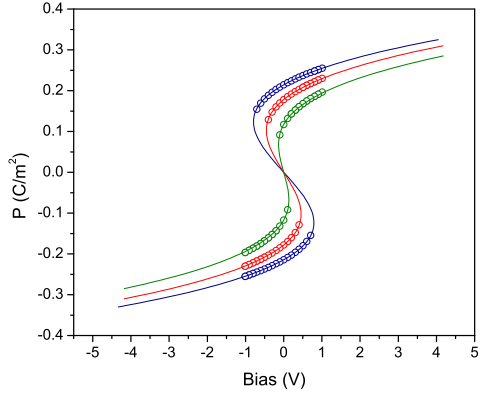}}
\caption{Theoretical $P-V$ plots for fully depleted films with  dopant concentrations $N_D=10^{18}, 10^{19}$and $1.5\times 10^{19}$cm$^{-3}$ and the curves calculated using equation~(\ref{cubic:OK}) with $\xi=0$ and $\alpha '$ and $\beta '$ extracted using the simulated $\epsilon(T)$ and $P_s(T)$.}
\label{FullDeplFits:OK}
\end{figure}
Provided the effect of strain is to simply rescale the LGD coefficients then if these modified  $\alpha '$ and $\beta '$ are extracted from zero-field measurements as discussed above, one expects that the hysteresis loop should fit the  cubic equation \cite{Catalan}
\begin{equation}
\alpha 'P + \beta ' P^3= E + \xi
\label{cubic:OK}
\end{equation}
where $\xi$ is a constant and $E=V/d$. But what happens when we have space charge?

We have used our model with $C=4.44\times 10^5$K and $T_c=548$K to simulate the temperature dependence of $P_s$ and $\chi$, where $\chi$ was calculated from polarization values at the tickle voltage of 50mV (as it would be experimentally) using
\[\chi=\frac{P(50\textrm{mV})-P(-50\textrm{mV})}{100\textrm{mV}}d.\]
For  partial depletion two cases were considered: 1) the applied 50mV drops entirely across the reverse biased contact i.e. $f_f=0$, and 2) half the applied bias drops across each contact i.e. $f_f=\frac{1}{2}$;  the difference between the extracted Curie constants for the two cases was negligible. In what follows, $\alpha^\prime$ and $\beta^\prime$ are coefficients that would be extracted experimentally from $\epsilon(T)$ and $P_r(T)$ measurements, modified from the bulk values $\alpha$ and $\beta$ by strain and space charge.
\begin{figure}[h]
\centering
\scalebox{1.0}{\includegraphics{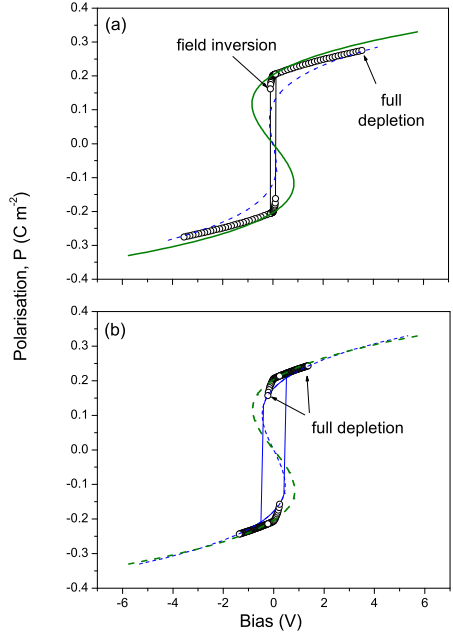}}
\caption{Theoretical $P-V$ plots for partially depleted films with $V_{bi}=0.2$V and (a) $N_D=1.5\times 10^{19}$cm$^{-3}$ and (b) $N_D=10^{19}$cm$^{-3}$ are shown as open cirles. The thick solid  curves were calculated using equation~(\ref{cubic:OK}) with $\xi=0$ and $\alpha '$ and $\beta '$ extracted using the simulated $\epsilon(T)$ and $P_s(T)$. Curves expected for fully depleted films with same $N_D$ are also shown for comparison (dashed).}
\label{PartDeplFits:OK}
\end{figure}

From figure~\ref{FullDeplFits:OK} we can see that in fully depleted films the experimentally extracted $\alpha '$ and $\beta '$, when substituted into  equation~(\ref{cubic:OK}), are expected to reproduce the hysteresis loops perfectly. This is not the case for partially depleted films. Figure~\ref{PartDeplFits:OK}(a) shows that the experimentally extracted parameters do not give a good fit to the loop for all but very low fields; in fact the loops are not well described by the cubic (\ref{cubic:OK}) even if we allow all the parameters to float in the fit. Decreasing the dopant concentration but keeping the built-in bias constant at 0.2V decreases the voltage range  over which the film is partially depleted (figure~\ref{PartDeplFits:OK}(b)). The hysteresis loop starts to resemble closer that of a fully depleted loop and the fit using experimentally extracted $\alpha '$ and $\beta '$ coefficients is expected to be better.

The extent of depletion in ferroelectrics has been a subject of debates for many years (for a discussion see chapter 5 of \cite{JimsBook}), with some authors believing them to be almost always fully depleted due to their insulating nature, and others  arguing them to be sometimes partially depleted due to the semiconducting behaviour often observed in thin films. Extracting  the LGD coefficients using zero field dielectric and polarization measurements and looking at how well they recreate the observed hysteresis loops can in principle give a qualitative estimate of the extent of depletion in a film. In practice, other factors can affect the shape of the hysteresis loops. Parasitic series capacitances can cause a stretching on the loop along the field axis, increasing the apparent coercive fields. Morozovska \emph{et al.} also showed that free carriers can result in tilting of the loops, which is likely to be significant in the case of semiconducting films. Thus, using the procedure described here, a good fit might indicate full depletion, whereas a poor fit may be caused by factors other than partial depletion.

\begin{figure}[h]
\centering
\scalebox{0.4}{\includegraphics{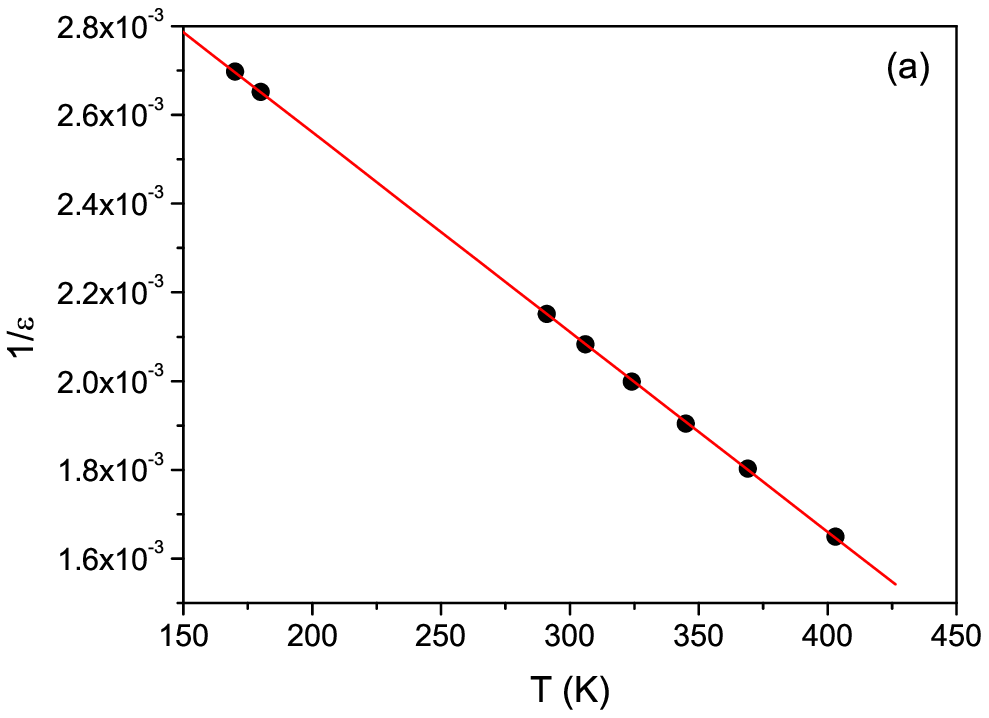}
\includegraphics{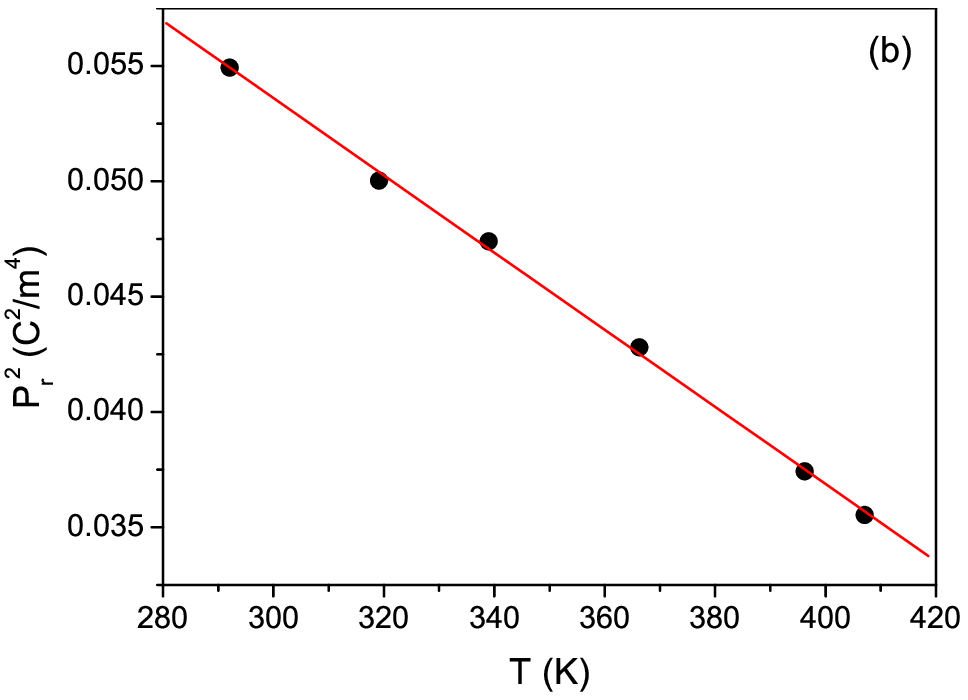}}
\scalebox{0.8}{\includegraphics{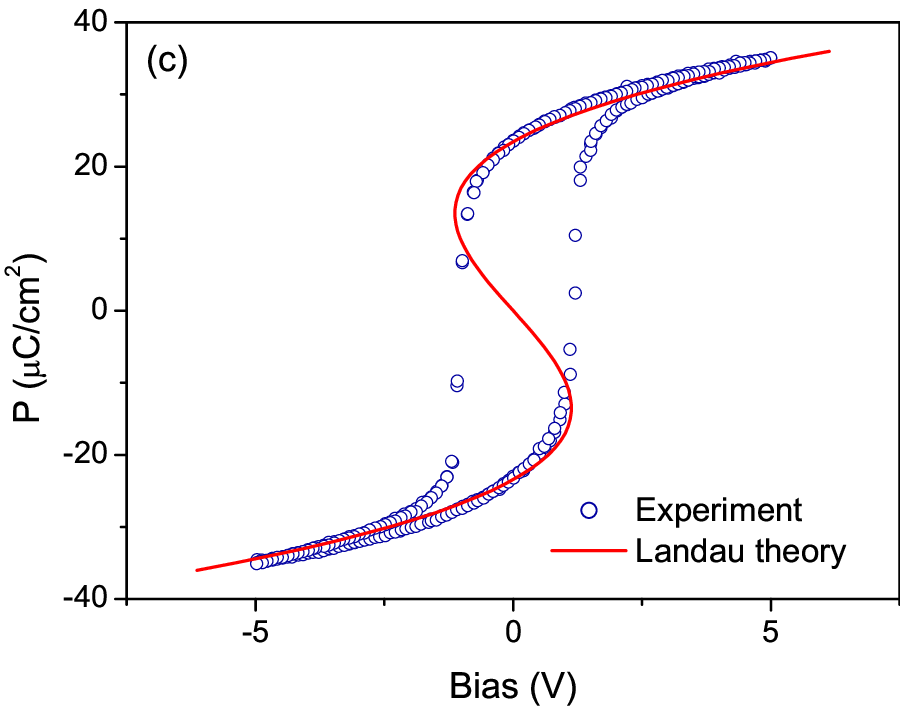}}
\caption{Temperature dependences of (a) $\epsilon^{-1}$, (b) $P_r^2$, and (c) LGD fit to the 1kHz hysteresis loop using  $\alpha '$ and $\beta '$ extracted from (a) and (b).}
\label{Experimental:OK}
\end{figure}

As an example, we present some experimental measurements on 144nm thick PbZr$_{0.4}$Ti$_{0.6}$O$_3$ capacitors from Samsung. The films deposted by a sol-gel method from 99.9999\% purity targets have a columnar structure on top of the [111] oriented Pt bottom electrodes, providing almost epitaxial interfaces  for individual grains. Together with the Ir/IrO$_2$ top electrodes the structures are fully integrated and embedded in Si. Further details of the samples and their full electrical characterisarion will be published elsewhere, here we present only the temperature dependence of the capacitance measured with a 50mV tickle bias using the HP4192A impedance analyser (figure~\ref{Experimental:OK}(a)) and the remanent polarization  obtained using the Radiant Precision Pro tester (figure~\ref{Experimental:OK}(b)). From the plot of $\epsilon^{-1}(T)$ the Curie constant was calculated to be $4.44\times 10^5$K; $P_r^2(T)$ plot yielded a value for $\beta '$ of $1.52\times 10^9$m$^5$C$^{-2}$F$^{-1}$ and $T_c=620$K. A room temperature hysteresis loop is shown in figure~\ref{Experimental:OK}(c) together with the LGD curve (\ref{cubic:OK}) using the calculated parameters (and $\xi=0$). The excellent fit over the whole range of biases investigated indicates that the material is probably fully depleted \cite{NoteOn111}. This is consistent with the large resistivity expected and observed in these high purity samples, as well as with the leakage mechnism  believed to be due  bulk-limiting Poole-Frenkel conduction (the current-voltage data will be published separately). 

\begin{figure}[h]
\centering
\scalebox{0.85}{\includegraphics{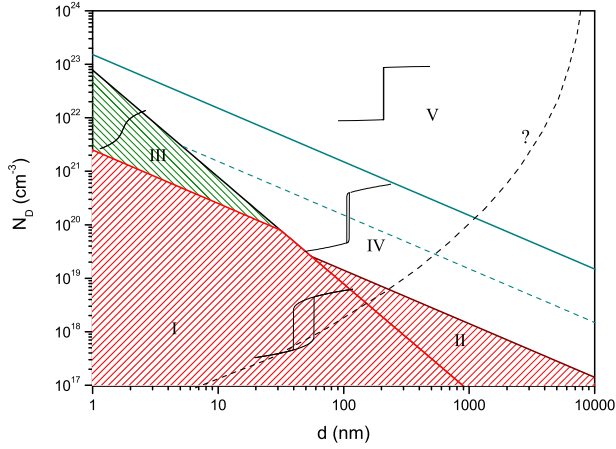}}
\caption{Regions of full (shaded) and partial depletion (unshaded) assuming $V_{bi}=0.1$V, $P_r=20\mu$C/cm$^2$ and $\epsilon=890$.}
\label{ND_vs_d:OK}
\end{figure}

So far we have only discused how the extent of depletion affects the shape of hysteresis loops without mentioning the conditions necessary for full or partial depletion to occur. Figure~\ref{ND_vs_d:OK} shows regions of full (shaded) and partial depletion as a function of donor impurity concentration and film thickness. For each region the expected shape of the hysteresis loop is also sketched. Samples with a built-in voltage $V_{bi}$  and with a dielectric constant of $\epsilon$ will be fully depleted at zero applied bias if 
\begin{equation}
N_D\apprle \frac{8\epsilon\epsilon_0 V_{bi}}{qd^2}
\end{equation}
giving the line with a slope of $-2$ on the $\log N_D$ vs. $\log d$ plot in figure~\ref{ND_vs_d:OK}. Within the fully depleted region the $P-E$ loops should be close to the ``ideal'' loop provided $N_D$ is below a certain value. Above this value space charge effects become important and the $P-E$ loops will rapidly shrink over  a narrow range of $N_D$ until the $P_r$ is almost zero. This transition happens when the field due to space charge becomes comparable to the depolarising field i.e.
\begin{equation}
N_D\approx \frac{2P_r}{q\epsilon d}
\end{equation}
giving the boundary which separates the loops with almost ``ideal''  from those with negligible $P_r$ (regions I and III respectively). We should note, that all boundaries in figure~\ref{ND_vs_d:OK} are not sharp, with a continuous but rapid transition between ``ideal'' and zero $P_r$ and $E_c$.

As depletion widths increase with applied bias a partially depleted sample may become fully depleted before the coercive field is reached. In such cases (region II) the hysteresis loop will still be similar to that of a sample which is fully depleted at zero bias. In the unshaded regions (IV and V) samples are expected to be partially depleted for most biases of interest. Within region IV the $P-E$ loops should become slimmer with increasing $N_D$. Across the two blue lines (dashed and solid) the coercive fields  drop to 5\% and $<1$\% respectively of the ``ideal'' value such that in region V the coercive field is expected to be negligible. It may seem that regions IV and V are very desirable for applications due to the reduction of $E_c$ without loss of $P_r$, however in practice  increasing $N_D$ also increases the free carrier concentrations inside the film and thus the dielectric loss which leads to tilting of the loops as shown by Morozovska, as well as to higher leakage currents which bring their own detrimental effects. Therefore, experimentally the loops observed in region IV might be either leaky or slim and tilted.
At this point we should note that our assumption of neglecting the field in the space charge free regions of the partially depleted samples becomes invalid when the thickness becomes large enough that the resistivity of the bulk becomes comparable to that of the depletion layers thus the arguments above will become invalid for thick films beyond the dashed curve in figure~\ref{ND_vs_d:OK}. The position and shape of this curve is difficult to calculate (hence the question mark) and in fact is probably very sample specific. For thicker films the uniform doping approximation is also not valid as the dopant concentration may vary by several orders of magnitude from interface to bulk.

\begin{figure}[h]
\begin{center}
\scalebox{0.8}{\includegraphics{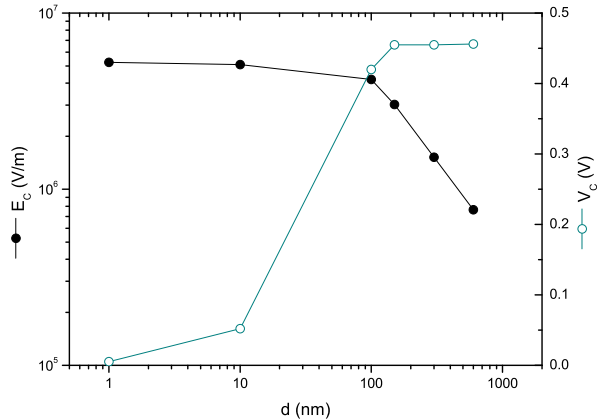}}
\end{center} 
\caption{Predicted thickness dependence of coercive field and voltage for $N_D=10^{19}$cm$^{-3}$.  The critical thickness at which the transition from full to partial depletion occurs will depend on the dopant concentration.}
\label{Ec-vs_d:OK}
\end{figure}

The boundaries in figure~\ref{ND_vs_d:OK} are only approximate, depending on sample-specific parameters like $V_{bi}$, $P_r$ and $\epsilon$, however, they should give a useful consistency check when interpreting experimental data. For example it makes no sense to interpret current-voltage data in terms of a partially depleted Schottky model if the hystersis loops are well fitted by the cubic equation indicating full depletion; alternatively strongly distorted $P-E$ loops indicate that the field distribution is probably not uniform inside the film. In their  recent work \cite{PintilOptoElec}, Pintilie \emph{et al.} have extracted a space charge concentration of $1.8\times 10^{19}$cm$^{-3}$ for a 250nm PZT film. In figure~\ref{ND_vs_d:OK} this falls into the partially depleted region IV, consistent with the authors' interpretation of the leakage mechanism.

It is also interesting to look at the thickness dependence of the coercive field for a particular dopant concentration. Well within region I the coercive field should be thickness independent as shown in figure~\ref{Ec-vs_d:OK}. It is then expected to decrease as the boundary is crossed and the partially depleted region IV is entered. Within this region polarization revesal is initiated by the strong fields within the depletion widths which are thickness indendent and thus it is the coercive voltage and not field which is expected to be constant. Examples of coercive field saturation for low $d$ include the Langmuir-Blodgett polymer film data of Ducharme \cite{Ducharme} (the interpretation of which, however, has been scrutinised by many researchers) and those of Jo \emph{et al.} on ultrathin BaTiO$_3$, which show a constant coercive field down to a thickness of 5nm. Thickness independent coercive voltages were reported by Trithaveesak \emph{et al.} \cite{Trithaveesak} for 100-400nm epitaxial BaTiO$_3$ films and by Kawashima and Ishiwara \cite{Kawashima} for 40-65nm SrBi$_2$Ta$_2$O$_9$ (SBT). Recently Pintilie and Alexe also observed coercive voltages independent of thickness (to within 10\%) in their epitaxial SrRuO$_3$/PZT/SrRuO$_3$ structures with $d$ in the range 	50--270nm \cite{PintilAlexe}.  For thicker films, when the bulk resistivity $\rho_b$ becomes important, switching still requires a constant coercive voltage $V_c$ to be dropped across the depletion layer, but now the measured coercive field will be given by
\[E_c=V_c\left(\frac{1}{d}+\frac{\rho_b}{R_d}\right)\]
where $R_d$ is the depletion layer resistance. This behaviour has been frequently observed for thicker films e.g. by Sidorkin \cite{Sidorkin} in lead titanate and  by  Merz in 1956 in BaTiO$_3$ \cite{Merz}. Merz argued that the $1/d$ behaviour was caused by a voltage drop across a surface layer which we can identify with a space charge depletion layer as was suggested even earlier by K\"{a}nzig~\cite{Kanzig}.

Finally we should mention that the highly non-linear field distributions expected for higher dopant concetrations have certain implications for conduction in ferroelectrics. In the case of fully depleted films the appearance of non-linearity in $E(z)$ coincides with the disappearance of the measurable polarization $\bar{P}$ and thus one is not likely to observe any strange conduction bahaviour if these films exhibit ferroelectric $P-E$ loops. However, this is not the case for partially depleted films. For capacitors with non-zero $V_{bi}$ the two depletion widths will not be equal even at zero applied bias,  as the field will be parallel to the polarization in one of them and antiparallel in the other. The depletion region with $E$ antiparallel to $P$ will be wider and the maximum field inside it will be lower. As the external field is applied to increase the voltage drop across this depletion region (i.e. the applied field is trying to switch the film) the magnitude of the field near the interface may actually \emph{decrease} resulting in possible negative differential resistivity if the usual Schottky thermionic emission mechanism is responsible for the current flow. Upon exceeding the coercive field and switching of the sample, the magnitude of the field at the interface will increase discontinuously resulting in a large increase in current. The large difference between switching and non-switching currents has been observed previously by Bloom \emph{et al} \cite{Bloom}, whereas the negative resistivity is only expected for a certain quite narrow range of $N_D$ and thus may not be so common. The large sudden change in resistivity near the coercive field can  be exploited in resistive switching devices.

\section{Discussion}
Having presented many interesting results we should now review the assumptions of our model and discuss its limitations. Let us start with the gradient term in equation~(\ref{FreeEnergyBulk:OK}) which we have chosen to neglect in this work. To our knowledge, the coefficient $\kappa$ has not been measured directly. Perhaps the only value so far was obtained by fitting the thickness dependence of the Curie temperature to a thermodynamic model, giving  $\kappa=4.5\times 10^{-10}$m$^3$F$^{-1}$ for lead titanate \cite{kappa}. We have used this value to calculate the relative contribution of the gradient term to the free energy and found that even for the highest $N_D$ considered in our calculations the gradient term amounted to only about 1\% of the total energy and thus we were justified to ignore it. However, we also ignored surface effects which may be quite considerable for very thin films and therefore our description may not be valid for very low $d$. Surface effects have been previously considered by several researchers but estimation of their significance is still very difficult due to lack of experimental values for parameters such as $\kappa$ and $\lambda$. 

Another assumption of our model was that the film is in a monodomain state. It was argued by Brennan \cite{Brennan} that the more stable state may be achieved by relaxation of the polarization to form two domains with polarization vectors pointing head to head. However, it was also argued that such a relaxation would require enough charge to be injected to compensate the highly charged domain boundary. We have not investigated the possibility of domain formation and this should be the subject of future work.

Finally, we should say a few words about coercive fields. One major drawback of using Landau theory for ferroelectrics is that it assumes homogeneous switching and predicts coercive fields orders of magnitude higher than observed experimentally. Moreover, nucleation is known to occur near the interface. The dielectric constant and polarization, however, are  well described by the thermodynamic approach. We show that the experimentally observed low coercive fields can be consistent with Landau theory predictions provided space charge, which is always present in reality, is accounted for.

Nucleation is often observed at specific sites near the interface and many models have been proposed to try and undestand how switching occurs. One role of defects or nucleation sites was discussed by Ishibashi. In his model, nuclei with reversed polarization are always present. The energy barrier for unswitched dipoles is reduced by their interaction with the reversed dipoles leading to a reduced coercive field, now dictated by the energy required to move a domain wall. Another possibility is the enhancement of local fields by charged defects. If this local field exceeds the coercive field a nucleation site will be created. This is precisely the effect of space charge. Even for uniformly doped films the field at the interface can be  orders of magnitude larger than   that in the bulk or the measured $E=V/d$, thus as far as  nucleation is concerned, space charge acts as an extended defect at the interface. As was shown, even moderate dopant concentrations of about $10^{19}$cm$^{-3}$ can lower the measured  coercive field by orders of magnitude compared to the intrinsic LGD value. Local space charge fluctuations may be responsible for the observed preferential nucleation at specific sites rather than homogeneously across the whole interface. 

Whatever the nature of nucleation, domain wall motion or in fact the whole switching process, it is clear that the coercive field will, to a large extent, be determined by the local field near the site where switching begins and this in turn is obviously  heavily influenced by space charge. Thus the trends for coercive fields discussed above should be useful at least qualitatively.


\section{Conclusions}
We have used the phenomenological Landau-Ginzburg-Devonshire theory to investigate numerically the effects of inhomogeneous field distributions due to stationary space charges on the ferroelectric properties of ferroelectric films. Cases where the samples are fully and partially depleted were considered. For fully depleted films both, polarization and coercive field values were found to decrease with increasing impurity concentration, as previously found by other authors. The field profile in  samples with high $N_D$ was found to have a cubic-like spatial distribution, unlike the linear profile expected for semiconductors; this limits the range of validity of usual semiconductor models for describing ferroelectrics. Overall the effect of space charge was to rescale the LGD coefficients, with the general shape of the $P-E$ loops still described by the usual cubic function.

In the case of partially depleted films, the remanent polarization depends strongly on the value of the built-in bias. The $P-E$ loops were found to be more square than the ideal space-charge-free loops or the fully depleted loops. A method was proposed which could help identify fully depleted films from temperature dependent capacitance and hysteresis measurements alone; this may be of particular interest for commercial device characterisation, where the physical parameters such as the ferroelectric thickness or composition are fixed.

The unusual field distributions expected for samples with high space charge concentrations may give rise to interesting current-voltage characteristics such as negative differential resistivity.

\section{Acknowledgements}
The authors are very grateful to Lucian Pintilie and Marin Alexe for their help in obtaining the low temperature capacitance data and would like to thank Gustau Catalan and Finlay Morrison for many fruitful discussions.


\begin{thebibliography}{300}
\bibitem{NoteOn111} Note, the validity of this model for [111] oriented samples, which may be the case here, is subject to certain conditions. It is still to be seen experimentally whether our samples satisfy these conditions, but is not the subject of this paper. Possible domain strtucture was also ignored.
\end{thebibliography}
\end{document}